# Quasiparticle Properties Below Coherence Onset in YbAl$_3$ Nanostructures

Dale T. Lowder[1], Gage Eichman[1], Yuxin Wan[1], Karthik Rao[1,11], Ruiwen Xie[2], Hongbin Zhang[2], Debjoty Paul[3], Shouvik Chatterjee[3], Darrell G. Schlom[4,5,6], Kyle Shen[5,7], Emilia Morosan[1,8,11], Douglas Natelson[1,8,9,10,11]

[1] *Department of Physics and Astronomy, Rice University, Houston, TX 77005, USA*
[2] *Institute of Materials Science, Technical University of Darmstadt, 64287, Darmstadt, Germany*
[3] *Department of Condensed Matter Physics & Materials Science, Tata Institute of Fundamental Research, Mumbai 400005, India*
[4] *Department of Materials Science and Engineering, Cornell University, Ithaca, New York 14853, USA*
[5] *Kavli Institute at Cornell for Nanoscale Science, Cornell University, Ithaca, New York 14853, USA*
[6] *Leibniz-Institut für Kristallzüchtung, Max-Born-Str. 2, Berlin 12489, Germany*
[7] *Laboratory of Atomic and Solid State Physics, Department of Physics, Cornell University, Ithaca, New York 14853, USA*
[8] *Smalley-Curl Institute, Rice Advanced Materials Institute, Rice University, Houston, TX 77005, USA*
[9] *Department of Electrical and Computer Engineering, Rice University, Houston, TX 77005, USA*
[10] *Department of Materials Science and NanoEngineering, Rice University, Houston, TX 77005, USA*
[11] *Rice Center for Quantum Materials (RCQM), Rice University, Houston, TX 77005, USA*

Mesoscopic transport measurements are underexplored as probes of quasiparticles and their properties in correlated metals. The mixed valence compound YbAl$_3$ exhibits a single-ion Kondo temperature of 670 K, while thermodynamic and transport properties (probed with specific heat, magnetic susceptibility, Hall effect, and resistivity) imply the onset of coherence of heavy fermion quasiparticles at $T^* \approx 37$ K. To characterize these quasiparticles, we utilize mesoscopic techniques familiar from weakly correlated conductors. In lithographically-defined nanowires etched from epitaxial films, we observe weak antilocalization magnetoresistance and universal conductance fluctuations, consistent with electronic coherence lengths of tens of nanometers. Additionally, analysis of Johnson-Nyquist noise measurements as a function of bias current reveal, within the context of a range of accepted models, a significant electron-phonon energy loss that increases with decreasing temperature, a finding that we contextualize within the broader properties of YbAl$_3$.

Mesoscopic electronic transport measurements, originally applied to weakly correlated metals and semiconductor heterostructures, allow experimental access to properties of itinerant charge carriers that are otherwise difficult to probe. Quantum corrections to semiclassical electronic conduction such as weak localization (WL) and universal conductance fluctuations (UCF) enable a detailed analysis of coherence lengths for quasiparticles [1,2], which can provide insight into materials with complex electronic behaviors.

YbAl3 is one such material and is a mixed valence compound with a single ion Kondo temperature of 670 K [3]. Measurements of magnetic susceptibility, magnetization, Hall effect, and magnetoresistance indirectly point to the onset of a heavy Fermi liquid below 40 K [3], and resistivity as a function of temperature is consistent with a crossover to a Fermi liquid $\rho = \rho_0 + AT^2$ below $T^* \approx 37\,K$ [4]. This is interpreted as the onset of fully well-defined, coherent heavy fermion quasiparticles. Specific heat and optical conductivity measurements [5] lead to an estimate of the effective mass $m^* \approx 25 - 30\,m_e$, where $m_e$ is the free electron mass. While the $T^2$ contribution to the resistivity is expected for Fermi liquid quasiparticles, to date there have been no reports directly measuring coherence effects or quantifying the electronic coherence length in this or other heavy fermion compounds.

In addition to the complex electronic behavior, YbAl3 has unusual coupling between the electrons and the lattice. Density functional and Boltzmann calculations show that the phonon drag contribution to the Seebeck response, due to momentum transfer between phonons and conduction electrons, is strong in this material, exceeding the purely-electronic band contribution [6]. Inelastic neutron scattering measurements [7] show strong coupling between the *f*-electron resonance and optical phonons near 32-35 meV associated with the aluminum atoms. At low temperatures, YbAl3 has a negative thermal expansion coefficient not seen in the nonmagnetic analog LuAl3 [8], with strong temperature variation below 20 K. Theory suggests [9] that this originates with continued renormalization of both phonon frequencies and the electron-phonon (e-ph) coupling as *f*-electron hybridization with the conduction electrons continues to evolve even below $T^*$.

In this paper, we present electronic transport and noise measurements in mesoscopic wires fabricated from epitaxial films of YbAl3. Our magnetotransport measurements reveal the presence of both weak antilocalization (WAL) and UCF, with these effects emerging at temperatures significantly below the coherence threshold $T^*$, as inferred from the temperature dependence of the resistivity $\rho(T)$. The field scale and magnitude of the WAL and UCF allow us to estimate the phase coherence length $L_\phi(T)$. Additionally, when analyzed in terms of accepted models, the Johnson-Nyquist noise in YbAl$_3$ wires shows a notable increase in e-ph energy loss as temperature decreases from 20 K to 3 K, as indicated by its dependence on the bias current.

~~Resistance~~ The residual resistivity of the YbAl$_3$ material is affected by the patterning process (see Table S1 for parameters for each of the wires measured). However, the temperature dependence of the resistivity remains similar to that of the unpatterned film. Fig. 1a shows the change in resistivity as a function of temperature for a representative wire 29 μm in length and 168 nm in width and for the unpatterned film. As with the unpatterned film, below ~ 37 K the resistance scales as $(L/wt)(\rho_{0,w} + A_w T^2)$, where $L$ is the wire length, $w$ is the wire width, and $t$, the wire thickness, is assumed to be 20 nm, with the resistance near 3 K being 9400 Ω. The coefficient $A_w = 1.9 \times 10^{-3} \pm 0.1 \times 10^{-3}$ μΩ·cm/K$^2$. The residual resistivity of this wire is $\rho_{0,w} \sim 103.6$ μΩ·cm, higher than the 48.5 μΩ·cm unpatterned film value. The residual resistivity varies from device to device due to the devices being processed sequentially at different times on the same substrate and variations in disorder due to fabrication process. A roughly temperature-independent, approximately linear-in-$B$ background is observed in the wire magnetoresistance MR= $\frac{(R(B)-R(B=0))}{R(B=0)}$ below ~ 20 K in Fig. 1b, which is also consistent with measurements done on bulk crystals [3,10]. This background is discussed further in the Supplementary Material (SM) in the context of the 4$f$ electronic structure.

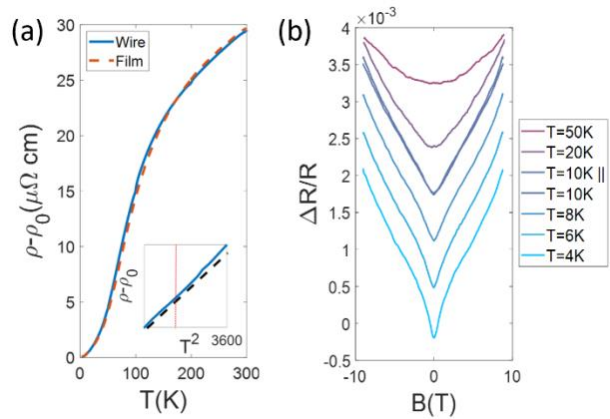

FIG. 1 (color online). (a) Change in resistivity ~~Resistivity~~ of an unpatterned film and a 29 μm long × ~ 160 nm wide wire of YbAl$_3$ showing that the fabrication of the nanowire does not significantly affect the properties of the material beyond changing the residual resistivity. Inset shows comparison of change in wire resistivity to $T^2$ contribution with a red dotted line indicating ~37 K. (b) Magnetoresistance data with field normal to the film (except where indicated). Curves are offset vertically for clarity. Measurement current was 200 nA.

Figure 2a shows the two-terminal magnetoconductance of a 168 nm wide wire measured with magnetic field directed perpendicular to the plane of the device. Measurement current was 200 nA, lowered to 100 nA for the lowest temperature to avoid self-heating. Wire width was found via scanning electron microscopy. Removing the linear-in-$B$ magnetoconductance background (linearly fitting the lowest temperature data at the fields higher than the onset of the cusp at $|B| = 2$ T up to 14 T and subtracting the slope acquired: $G(B) = G_{raw}(B) - slope * B$) reveals a peak in the conductance centered near zero field; this feature is consistent with WAL, with antilocalization expected given the strong spin-orbit coupling in this system from the Yb. At higher fields, fluctuations are observed in the conductance as a function of $B$ that look random but retrace themselves on subsequent field sweeps (Fig 2c) and grow and sharpen as $T$ is decreased (Fig 2d). These features are consistent with UCF.

WAL can be analyzed in terms of theoretical expectations [11,12] using the appropriate expression in the 2D limit (Eq. (1)). A comparison between the inferred coherence length and the wire width and thickness allows a check for self-consistency of the assumption of quasi-2D quantum corrections to the conduction. Here $\psi$ is the digamma function.

$$\left.\frac{R(B)-R(0)}{R(0)}\right|_{2D} = \frac{e^2}{2\pi^2\hbar}\frac{R(0)w}{L}\left[\psi\left(\frac{1}{2}+\frac{1}{2}\frac{\hbar}{2e|B|L_\phi^2}\right) - \ln\left(\frac{1}{2}\frac{\hbar}{2e|B|L_\phi^2}\right)\right] \quad (1)$$

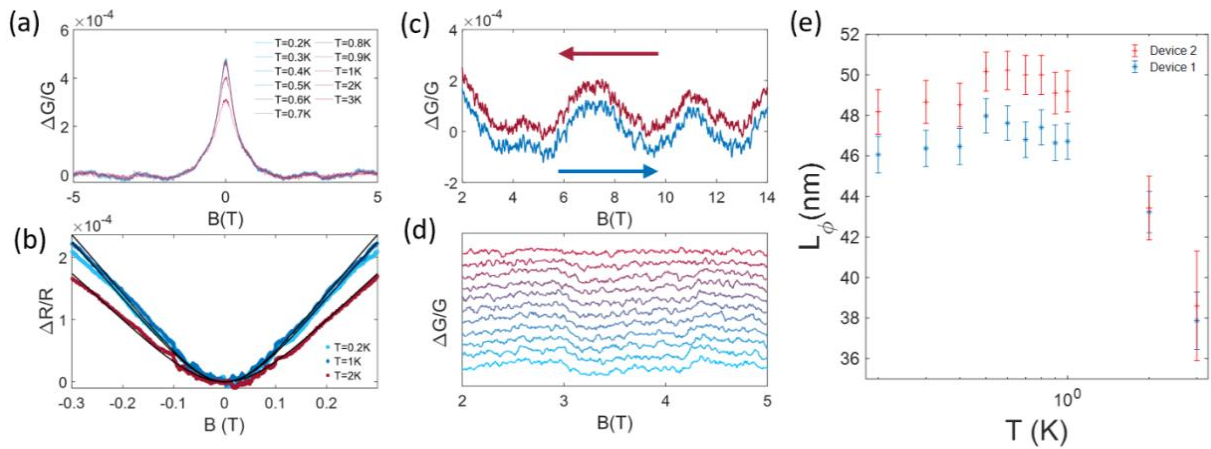

FIG. 2 (color online). (a) Magnetoconductance of a 168 nm wide device (device 1) with the linear-in-$B$ background subtracted, revealing WAL and UCF. (b) WAL magnetoresistance data with fits to Eq. (1) superimposed. (c) Magnetoconductance sweeps for device 2 at 0.2 K showcasing the retracing of the UCF signal from 2 T to 14 T and back. (d) Zoom in and vertical offset of the magnetoconductance data of device 1 between 2 T and 5 T showing the sharpening of the UCF signal as temperature is decreased. The color scale reflects the temperatures from panel (a). (e) Coherence length as a function of temperature as inferred from WAL data. Error bars are inferred from changing of $\chi^2$ as a function of the fit parameter $L_\phi$ at each substrate temperature.

Figure 2b shows the fits of the data with the fit range being limited ±0.3 T, to minimize the effects of UCF on the fitting results. The extracted coherence lengths $L_\phi(T)$ (Fig. 2e) are less than $w$ and greater than film thickness $t$, so the analysis in terms of quasi-2D WAL is internally consistent. The coherence lengths exceed the inferred elastic mean free path (~ 1.7 nm; see end material), also a requirement for the consistency of the usual derivation of the WAL correction. Apparent saturation of $L_\phi(T)$ at low temperatures is not unusual in general and can be a result of spin-flip scattering from paramagnetic impurities [13,14]. While pure stoichiometric YbAl3 should not contain such possible scattering centers, impurities or chemically distinct Yb-containing species at surfaces and (etched) boundaries could function as scattering sites. Further experiments would be needed to test this hypothesis, and this is beyond the scope of the present study.

In addition to quantifying $L_\phi$ via WAL, UCF can be seen at lower temperatures and can be used as another means of assessing coherence. The lowest temperature data ($T$= 0.2 K) is used, as it offered the strongest UCF signal. The WAL component between ±2 T is cut from the data, and the background is linearly fit between $|B| > 2$ T and $|B| < 14$ T and removed to give a flat average MR. The autocorrelation $F$ of $\delta G(B) \delta G(B + \Delta B)$ is taken (Fig. S2) to determine a correlation field $B_c$ such that $F(B_c) = \frac{1}{2} F(0)$ [15]. From $B_c$, $L_\phi$ can be determined [15] using Eq. (2) and Eq. (3) with $D$ being the diffusion constant calculated from data on single crystal YbAl3 [16,17]. The appropriate theoretical expression to use depends on how the thermal length scale $L_T = \sqrt{\hbar D/k_B T} \sim 29$ nm compares to $L_\phi$. When the two scales are comparable, $L_\phi$ exists in the crossover, and equations (2-3) below now define bounds for the value of $L_\phi$ via UCF with Eq. (2) giving $L_\phi \sim 37$ nm and Eq. 3 giving $L_\phi \sim 103$ nm. These values bracket the quantities found from WAL fitting, confirming that the coherence length is tens of nm in the YbAl3 wires.

$$L_\phi = \sqrt{0.49\frac{h}{e}\frac{1}{B_c}} \qquad\qquad L_T \ll L_\phi \qquad (2)$$

$$L_\phi = 1.14\frac{h}{e}\left(\frac{k_bT}{\hbar D}\right)^{\frac{1}{2}}\frac{1}{B_c} \qquad L_T \gg L_\phi \qquad (3)$$

The WAL and UCF data are consistent with well-defined Fermi liquid quasiparticles in this mixed valence heavy fermion system at temperatures well below the nominal onset of coherence, ~ 37 K. As expected for a Fermi liquid in the comparatively weakly disordered regime ($kF\ell \gg 1$), it is reasonable to consider the semiclassical picture of heavy charge carriers as diffusing through the material, but with quantum interference corrections as the complex phase of such a wavepacket remains well defined over distances of tens of nm. The dominant decoherence mechanism (e.g., electron-electron scattering; electron-phonon scattering at higher temperatures; extrinsic like spin-flip interactions with paramagnetic defects) in such a heavy fermion system can in principle be inferred from careful study of these mesoscopic corrections as a function of temperature and sample geometry.

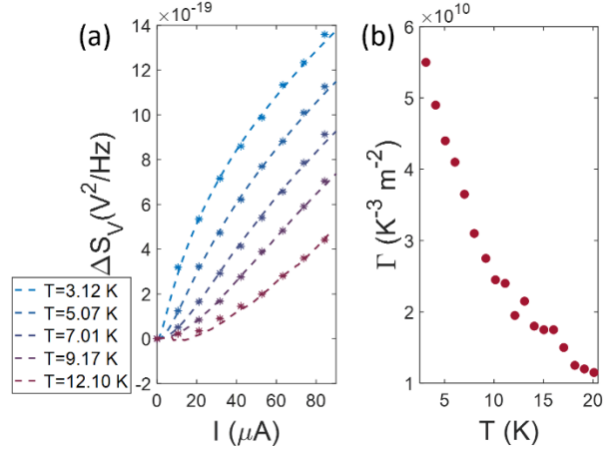

FIG 3 (color online). (a) Noise vs. bias current in a 265 nm wide and 27 μm long wire with fits to Eq. (4). (b) Calculated $\Gamma$ vs substrate temperature $T$ for same wire. Error bars (too small to resolve) are inferred from changing of $\chi^2$ as a function of the fit parameter $\Gamma$ at each substrate temperature.

In addition to quantum coherence corrections to conduction, the fabrication of YbAl$_3$ nanowires enables further measurements to quantify the inelastic scattering of the emergent heavy charge carriers. Noise measurements in nanowires as a function of bias current make it possible within a simple 1D thermal model to infer the energy loss rate from the conduction electrons to the phonons [18–20]. This approach has been employed previously in examining electron-phonon energy loss in mesoscopic AuCu structures [18] and gold nanowires [19,20].

$$\frac{\pi^2}{6}\frac{d^2 T_e^2}{dx^2} = -\left(\frac{eIR(T_e)}{k_B L}\right)^2 + \Gamma\left(T_e^5 - T_{ph}^5\right) \qquad (4)$$

In a Fermi liquid wire that is long compared to the distance scales associated with large energy transfer inelastic electron-electron scattering, at a given location the electronic quasiparticles are described by a Fermi-Dirac distribution with a local electronic temperature $T_e(x)$. In equilibrium, $T_e(x) = T_{ph}$, the uniform lattice temperature of the phonons. When the wire is driven with an applied voltage (described by the current, $I$, and the resistance, $R(T_e(I))$, at that voltage in terms of the $T_e$), Eq. (4) describes $T_e(x)$ along the wire, assuming that the phonons and the large contact pads are held at a fixed lattice temperature $T_{ph}$. Note that the differential resistance under bias is used in Eq. (4), though in this case the wire is essentially Ohmic (see Fig. S3). For a wire much longer than the effective e-ph inelastic scattering length, in this model the dissipated power from Joule heating dominantly transfers to the lattice, as characterized by $\Gamma$, an e-ph energy loss parameter. In this limit, except near the wire ends, $T_e(x)$ is approximately uniform [21]. Using the resulting temperature distribution and computing the Johnson-Nyquist noise of the wire, it is possible to model the voltage noise, $S_V$, as a function of bias current. Measurements of this type on pure Au wires [19,20] can be fit at multiple low temperatures with a single $T$-independent value of $\Gamma \approx 5 \times 10^9$ K$^{-3}$ m$^{-2}$. Similarly, in long wires of the quantum critical heavy fermion material YbRh$_2$Si$_2$ (YRS) [21], a single value of $\Gamma \approx 9 \times 10^9$ K$^{-3}$ m$^{-2}$ does a reasonable job of fitting the data from 3 K to 7 K. The $T^5$ functional form is expected to be valid in the degenerate Fermi regime when $T \ll \theta_D$, the Debye temperature [22]. We note that $T^4$ and $T^6$ dependences are predicted under certain circumstances [23], and alternative models (other exponents or more general $T$-varying mechanisms) become increasingly challenging to distinguish at higher temperatures due to the smaller changes in $T_e$. See SM for further discussion.

We measured the voltage noise as a function of bias in identically fabricated YbAl$_3$ wires of width 265 nm (device 3) using a low frequency cross-correlation setup described elsewhere [21,24] (see SM). The results are shown in Fig. 3. The Debye temperature in YbAl$_3$ is found to be $\theta_D \approx 337$ K [25], and thus the phonon population is largely frozen out at low temperatures. Phonons can, however, still be excited via driven electrons. At each substrate temperature, we find that the voltage noise as a function of current can be modeled well using Eq. (4), with $\Gamma$ as an adjustable parameter. Unlike the situation in Au or YbRh$_2$Si$_2$, however, fitting the different substrate temperatures requires a temperature-dependent $\Gamma(T)$ considerably larger over the whole temperature range than the value for Au.

Within this model of electron-phonon energy loss, the comparatively large $\Gamma$ is a signature of strong e-ph coupling. The effective electron-phonon length inferred from this [19,20] is around 1 μm at 3 K, considerably longer than the coherence lengths determined by the WAL and UCF. This is not surprising, given that coherence can be limited by small-angle, small energy transfer inelastic scattering [1], while the length computed from $\Gamma$ involves large energy transfer processes. Studies of thermoelectric properties in YbAl$_3$ [6,26,27] have indicated that electron-phonon scattering with strong energy transfer is likely near room temperature, given that the phonon drag component of the Seebeck response exceeds the bare electronic contribution. At low temperatures, inelastic neutron scattering [7] indicates that an Al-based optical phonon mode can resonate with the Yb 4$f$ level, causing dynamic changes in 4$f$ hybridization. The strong temperature variation of $\Gamma(T)$ implies that significant e-ph interactions persist and evolve even at temperatures well below the coherence temperature for the heavy fermion quasiparticles. This temperature dependence of $\Gamma$ occurs over the same range in which the thermal expansion coefficient of YbAl$_3$ is observed to be negative and temperature dependent [8].

This coincidence suggests that a single underlying mechanism involving the evolving $f$-electron hybridization in the coherent heavy fermion Fermi liquid state drives changes to the lattice and the e-ph energy transfer processes. A strong connection between $f$-electron hybridization and lattice structure has long been known in Ce compounds in the context of the Kondo Volume Collapse [28,29]. In that case, the free energy of the system depends strongly on the hybridization because of competition between the spin entropy of the local moments and the elastic energy cost of lattice deformation. In Yb mixed valence compounds, changes in 4$f$ occupancy are strongly associated with corresponding changes in lattice volume, with the $f^{13}$ limit having smaller volume and the $f^{14}$ limit having larger volume [9]. Other mixed valence Kondo lattice systems similarly have negative thermal expansion coefficients associated with valency changes and affected by chemical disorder [30].

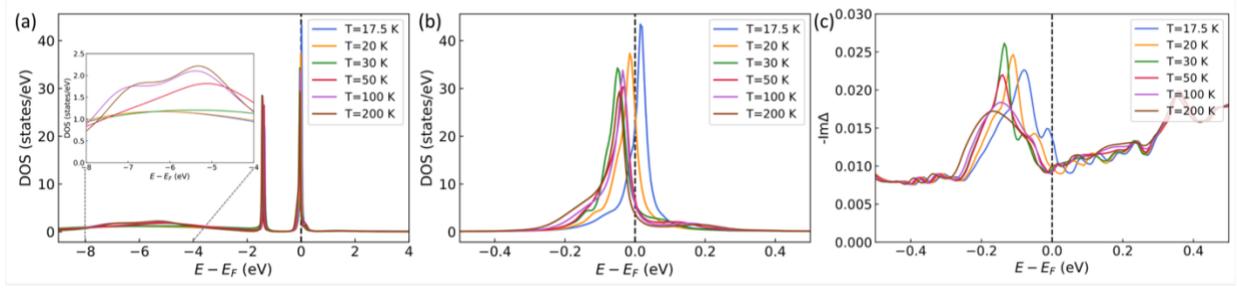

FIG. 4 (color online) (a) Calculated density of states (DOS) corresponding to Yb 4$f$ states at different temperatures. The inset zooms in the DOS at the range of [-8, -4] eV. (b) DOS of Yb 4$f$ states close to the Fermi level $E_F$ ([-0.5, 0.5] eV). (c) Hybridization function of Yb 4$f$ states in the energy range of [-0.5, 0.5] eV.

Directly examining the connection between evolving heavy fermion quasiparticles and lattice degrees of freedom is challenging. To gain insight, we have applied charge self-consistent density functional theory combined with dynamical mean-field theory (DFT+DMFT) calculations [31–33] to obtain the electronic structure of YbAl$_3$, which considers simultaneously the strongly localized 4$f$ electrons and the spin-orbit coupling (SOC). The phonon spectra and DOS of YbAl$_3$ were calculated with Quantum Espresso (QE) [34] using the DFPT method [35]. Following the idea proposed in Ref. [36] to address the electron-phonon coupling in correlated materials, we selected two representative distortion modes (as discussed below) and carried out the charge self-consistent DFT+DMFT calculations to obtain the self-energy difference at T = 20 and 200 K, respectively.

Figure 4 shows the temperature dependence of the Yb 4$f$ DOS and hybridization function. It can be seen from Fig. 4(a) that, with decreasing temperature, especially when T < 50 K, the coherence feature at the Fermi energy starts to develop, corresponding to the spectral weight transfer from the energy range of [-8, -4] eV to a range adjacent to $\epsilon_F$. By zooming in the Yb 4$f$ DOS around the $\epsilon_F$, we can clearly observe the increasing Yb 4$f$ DOS at the Fermi level with decreasing temperature (see Fig. 4(b)). In addition, the corresponding peak of the $J = 7/2$ states below the $\epsilon_F$ is shifted closer to the $\epsilon_F$ with lowering temperature, which has also been observed from the angle-resolved photoemission spectroscopy (ARPES) [37]. Furthermore, the hybridization function (Fig. 4(c)) that characterizes how 4$f$ electrons hybridize with the conduction electrons demonstrates a similar trend as that of the DOS, with stronger hybridization when $T$ is reduced below 30 K.

To investigate the temperature dependence of electron-coupling, we have calculated also the self-energy difference between the perturbed cell ($\Sigma^{ph}$) and the unperturbed cell ($\Sigma^{cn}$) at $T = 20$ and 200 K, respectively. Based on the phonon band structure, Yb and Al contribute mainly to the low-frequency and high-frequency regions, respectively. Accordingly, we selected two representative distortion modes (Fig. S2), i.e., high-frequency mode G and low-frequency mode R, for further study on the self-energy difference ($\Sigma^{ph} - \Sigma^{cn}$) corresponding to the $J = 7/2$ states introduced by the lattice distortion. For the imaginary part of $\Sigma^{ph} - \Sigma^{cn}$, the change caused by the lattice distortions seems more substantial at low temperature, especially for the mode G (Fig. S2). These changes and the growth in $g(\epsilon_F)$ are consistent with an increasing e-ph scattering parameter $\Gamma$ with decreasing $T$.

In summary, these mesoscopic transport and noise studies provide insights into the electronic excitations of this mixed valence compound. The observations of WAL and UCF at temperatures well below the coherence temperature $T^* \approx 37$ K are direct evidence for the existence of coherent heavy fermions. These mesoscopic phenomena give quantitative access to the phase coherence of the emergent quasiparticles. Both WAL and UCF analyses find $L_\phi$ of tens of nanometers. The very weak temperature dependence of $L_\phi$ implies that while the quasiparticles are coherent ($L_\phi \gg \ell$), inelastic processes strongly affect the coherence in these nanostructures, whether from extrinsic sources (e.g., paramagnetic impurities and defects) or from intrinsic sources (still evolving effective e-e scattering).

The Johnson-Nyquist noise measurements reveal an unusually strong and temperature-dependent e-ph coupling in YbAl$_3$, with the energy loss parameter $\Gamma$ increasing significantly as temperature decreases from 20 K to 3 K. This temperature dependence is coincident with the negative and $T$-dependent thermal expansion coefficient previously reported in bulk crystals, suggesting a common underlying mechanism likely related to the continued evolution of $f$-electron hybridization with conduction electrons below $T^*$. The magnitude of $\Gamma$ in YbAl$_3$ considerably exceeds that of Au and even other heavy fermion compounds like YbRh$_2$Si$_2$ by an order of magnitude, consistent with the exceptionally strong e-ph interactions theoretically predicted and inferred in other measurements.

These results demonstrate that mesoscopic transport techniques, traditionally applied to weakly correlated materials, can serve as powerful probes of complex electronic systems. By extending these approaches to strongly correlated materials like YbAl$_3$, we gain valuable insights into the nature of heavy fermion quasiparticles and their interactions with the lattice. Future work applying these techniques to other correlated electron systems may provide further understanding of the interplay between electronic correlations, coherence, and e-ph coupling in complex quantum materials.

**Acknowledgments**

DTL, YW, and DN acknowledge support for this research from DOE BES award DE-FG02-06ER46337, the nanofabrication infrastructure of the Rice Shared Equipment Authority. GE and DN acknowledge support from NSF DMR-2329111 for the development of the Kelvin double bridge used in the magnetoresistance measurements. KR and EM acknowledge support from the Robert A. Welch Foundation grant C-2114 for measurements performed in the dilution refrigerator. SC, DGS, and KS acknowledge support from NSF DMR-2104427 for the YbAl$_3$ film growth and characterization. RX and HZ gratefully acknowledge the computing time provided to them at the NHR Center NHR4CES at RWTH Aachen University (project number: p0024007). This is funded by the German Federal Ministry of Education and Research, and the state governments participating on the basis of the resolutions of the GWK for national high performance computing at universities (www.nhr-verein.de/unsere-partner).

**Data availability**

The data for this study are available via public archive. [38]

## Supplementary Material

***Sample fabrication.*** Epitaxial films of YbAl$_3$ are grown on MgO substrates, with the full stack being 5 nm LuAl$_3$, and 15 nm YbAl$_3$ according to the method in Ref. [1]. The Lu has a full 4$f$ shell and is thus magnetically inactive. Resistivity as a function of temperature for the film is shown in Fig. 1 and resembles that for bulk single crystals [2], though the film has a residual resistivity ratio (RRR) of about 1.5 while the single crystal has RRR ≈ 40. Resistivity below about 37 K scales as $\rho_{0,f} + A_f T^2$, as expected based on the single crystal results and as reported previously. The coefficient $A_f = 1.7 \times 10^{-3} \pm .1 \times 10^{-3}$ μΩ·cm/K$^2$ is found by fitting $\rho(T)$ at temperatures below 30 K. The residual resistivity of this film $\rho_{0,f} \approx 48.5$ μΩ·cm is considerably higher than both the reported bulk residual resistivity in single crystals [2] of 1.3 μΩ·cm, and the residual resistivity of 26 μΩ·cm reported in thicker films [1]. The measured $A_f$ is in the range between what has been reported in single crystals and hot-pressed samples [2] and is higher than that seen in thicker films. These observations are consistent with a trend toward higher $A_f$ with increasing disorder.

To fabricate nanowires, electron beam lithography is first used to pattern the contacts, which are made through deposition (5 nm Ti + 80 nm Au + 32 nm Cr) and subsequent liftoff. A hard etch mask for the nanowire itself (60 nm Ge + 40 nm Cr) is then similarly prepared using e-beam lithography, e-beam evaporation, and liftoff. Positive photoresist S1818 from Microposit is used to protect the rest of the chip during Ar Etching. An Oxford Plasmalab System 100/ICP 180 with Ar plasma, 14 mTorr and 120 Watts of RF power is employed to etch away approximately 35 nm of the 20 nm film with an etch rate of ~5.9 nm/min. The Cr is used as a hard mask to protect the contact pads and wire, with the Ge acting as a buffer layer to prevent damage to the wire. Test measurements show that residual Ge/GeO$_x$ left behind by this process is electrically insulating below 150 K and serves as a protective layer for the wire.

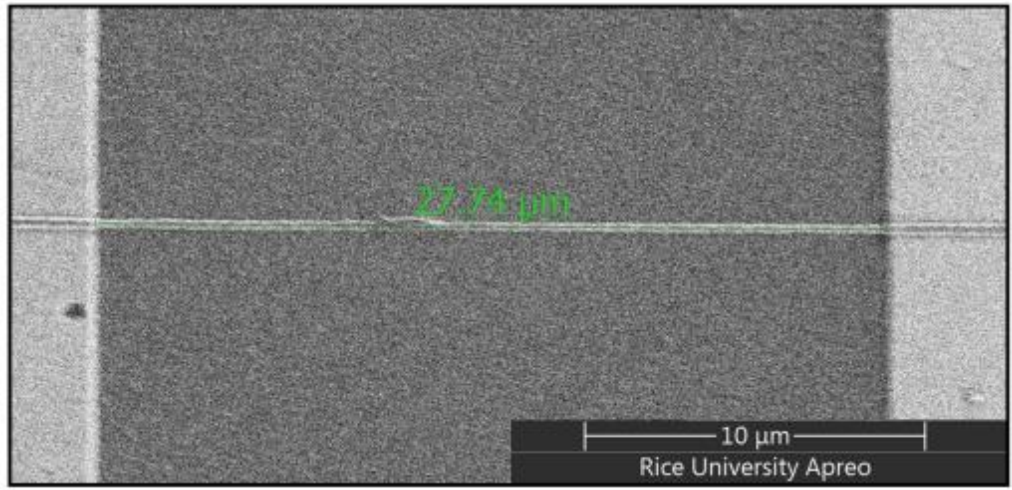

FIG S1. Scanning electron microscopy image of a representative wire used for measurement.

***Individual device resistivities.***

As mentioned in the main text, the patterning process affects the residual resistivities of the resulting wires, which were fabricated serially in different runs from material on the same substrate. Wire lateral dimensions were obtained from post-measurement electron microscopy.

| Device | Dimension (t=20 nm) | $\rho_0$ (μΩ·cm) |
|---|---|---|
| 1 (FIG 1) | 160 nm x 29 μm | 103.6 |
| 2 (FIG 2) | 168 nm x 28 μm | 110.4 |
| 3 (FIG 2) | 154 nm x 29 μm | 120.5 |
| 4 (FIG 3) | 265 nm x 28 μm | 55.4 |

TABLE S1. List of different devices used in this study and their dimensions and their residual resistivities.

*Noise measurement.* Noise spectroscopy measurements were done across 2 two-stage amplifier chains in accordance with the cross-correlation measurements technique (cite). The first stage amplifier consists of a LI-75 with approximate input noise of 1.2 nV/√Hz. The setup is calibrated to account for the large resistance of the samples by taking Johnson-Nyquist noise measurements of the sample at various temperatures and plotting them against the cryostat temperatures to obtain the actual gain of the system and any noise offset.

Samples are measured in a two-terminal configuration with a single noise measurement consisting of 300 averages of the cross-correlated signal. This measurement is repeated 3 times for each voltage $V$ at each temperature $T$. The spectra are averaged together to a single point across an observed large flat bandwidth to determine the noise power of the white noise signal at a given $V$ and $T$.

*Quantitative analysis of transport parameters.* Prior heat capacity measurements of bulk YbAl$_3$ have found [3] the Sommerfeld coefficient to be $\gamma$ = 45 mJ/K$^2$·mol. From the lattice constant of 4.20 Å, the volume coefficient of electronic specific heat is $\gamma_v$ = 1.01× 10$^3$ J/K$^2$m$^3$ = $(\pi^2/3)k_B^2 g(\epsilon_F)$, in a Sommerfeld picture, where $g(\epsilon_F) \approx 1.6 \times 10^{48}$ J$^{-1}$m$^{-3}$ is the density of states (DOS) at the Fermi level. From the Einstein relation $1/\rho_{0,f} = e^2 D g(\epsilon_F)$, the diffusion constant for the carriers in the film is $D$ = 5.0 × 10$^{-5}$ m$^2$/s. Measurements of deHaas-van Alphen oscillations in single crystals [4] have found an effective mass of $m^* \approx 23\ m_e$ for the carriers; within a Sommerfeld model. Combining this with $g(\epsilon_F)$ gives a Fermi velocity $v_F \approx 4.2 \times 10^4$ m/s, consistent with other reports [5]. Assuming 3D diffusion in the thin film, the elastic mean free path inferred from $D$, $v_F$, and the residual resistivity is $\ell_f \approx 3.5$ nm. Assuming no changes in the actual electronic structure of the material (and hence $g(\epsilon_F)$, $v_F$, etc.), the inferred diffusion constant in the wire is $D_w \approx 2.3 \times 10^{-5}$ m$^2$/s , implying a (residual) elastic mean free path in the etched wire of $\ell_w \approx 1.7$ nm.

*Autocorrelation and UCF analysis.* The autocorrelation is computed for fields between -14 T and -4 T for the device whose data is shown in Fig. 2c. An essentially identical result is found for computing the autocorrelation between 4 T and 14 T. The lower bound on field magnitude was chosen to mitigate any complications from the WAL feature near zero field.

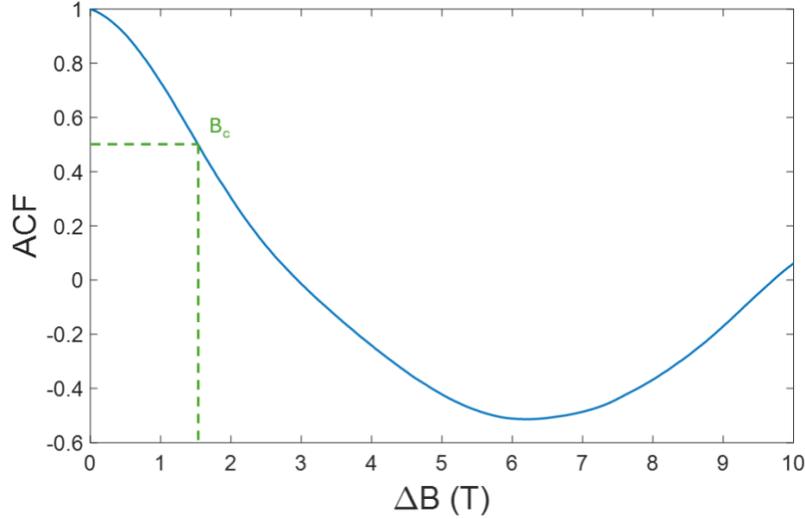

FIG S2. Autocorrelation of the magnetoconductance data for the device shown in Fig. 2c, for fields between -14 T and -4 T. Conventionally the field scale $B_c$ is set when the autocorrelation falls to ½.

*Noise measurement technique*. This procedure is essentially identical to that published previously[6]. The sample is mounted in a custom noise probe for insertion into a Quantum Design PPMS cryostat. The noise probe is electrically isolated from the PPMS. Bias current to the device is supplied via a heavily filtered National Instruments NI-DAQ6521 voltage source. The voltage across the device under test is measured by two parallel amplifier chains at room temperature, each consisting of a NF LI-75A voltage preamplifier (gain = 100) and a SR560 voltage preamplifier (gain = 100). The two voltage time series are sampled, with each sample consisting of 2 million data points with a sampling rate of 10 MHz, using a high-speed Picoscope 4262. The cross-correlation is computed between the sampled time series 300 times and the results are averaged. This is repeated 3× and the results are averaged for each datapoint. The cross-correlation approach is intended to minimize the effect of the input noise of the amplifier chains. The conversion between the measured noise and the true sample noise (accounting for amplifier gains, the cross-correlation calculation, and any remaining amplifier noise background) is calibrated by measuring the temperature-dependent Johnson-Nyquist noise of a resistor comparable to the sample resistance.

*Electron-phonon energy loss temperature dependence.* Figure 3b in the manuscript shows the inferred temperature dependence of the coefficient $\Gamma$ from Eq. (4). The $T^5$ behavior in the model of Eq 4 for energy transfer between the electrons and the phonons is expected to be valid when the temperature is well below the Debye temperature, and when the electron system is in the degenerate Fermi regime (which should be true in the present case below 37 K since the resistivity shows the characteristic $T^2$ dependence) [6] . In the case of YbAl$_3$ the Debye temperature, $\Theta_D \approx 337$ K [8], is well above all the temperatures used in this study. Two other power laws have been put forward as candidate dependences at low temperatures, $T^4$ and $T^6$, in the "dirty" limit when thermal phonon wavelengths are much longer than the electron elastic mean free path [7]. A $T^4$ scaling can appear when including quantum interference effects in the presence of static disorder, and a $T^6$ scaling can appear if vibrational dragging of the disorder potential is relevant.

As mentioned in the main text, the measured noise increase with bias, $\Delta S_V$, is a proxy for the increase of the (essentially uniform) electron temperature, $\Delta T_e$. As the substrate temperature is increased, particularly above 10 K, that temperature increase is relatively small, making it challenging to distinguish between alternative models for electron-phonon energy loss, or to identify whether the dominant electron-phonon cooling mechanism may somehow change at high temperatures.

We have tested alternative possibilities. First, in Fig. S3, we consider the exponents 4-6 assuming a single exponent and a single prefactor for each temperature. As discussed further below, the $T^5$ model fits all the data with minimal residuals (compared to the alternatives) at the lowest temperatures. The unusual conclusion that the coefficient of electron-phonon energy loss must grow strongly with decreasing temperature to match the data is true for all three models (Fig. S3d-f), albeit to different degrees, even if one limits consideration to substrate temperatures $\leq$ 12 K.

To further examine the possibilities, we also fit models with exponents ranging from 2-6 at a representative higher temperature (15 K) and at the lowest temperature (3 K). While visually the models yield similar maximum temperatures at 15 K due to the comparatively small change in $T_e$, the residuals at specific bias currents (Fig. S4) and variance across the whole bias range (Table S2) of the fits show that the $T^5$ model still appears to be the best fit. This conclusion is drawn from the minima in the variance at $n = 5$ and the similar scattering of all the models about the experimental data. Similarly, at 3 K, the residuals at specific bias currents (Fig. S5) and variance (Table S3) show clearly that the $T^5$ model is the best fit to the data.

While one cannot definitively rule out the possibility of some unexpected, more complex, temperature-dependent electron-phonon cooling mechanism at the highest temperatures, within the context of the accepted models (exponents 4-6), the conclusion that the coupling parameter for energy loss from electrons to the lattice grows with decreasing temperature is robust.

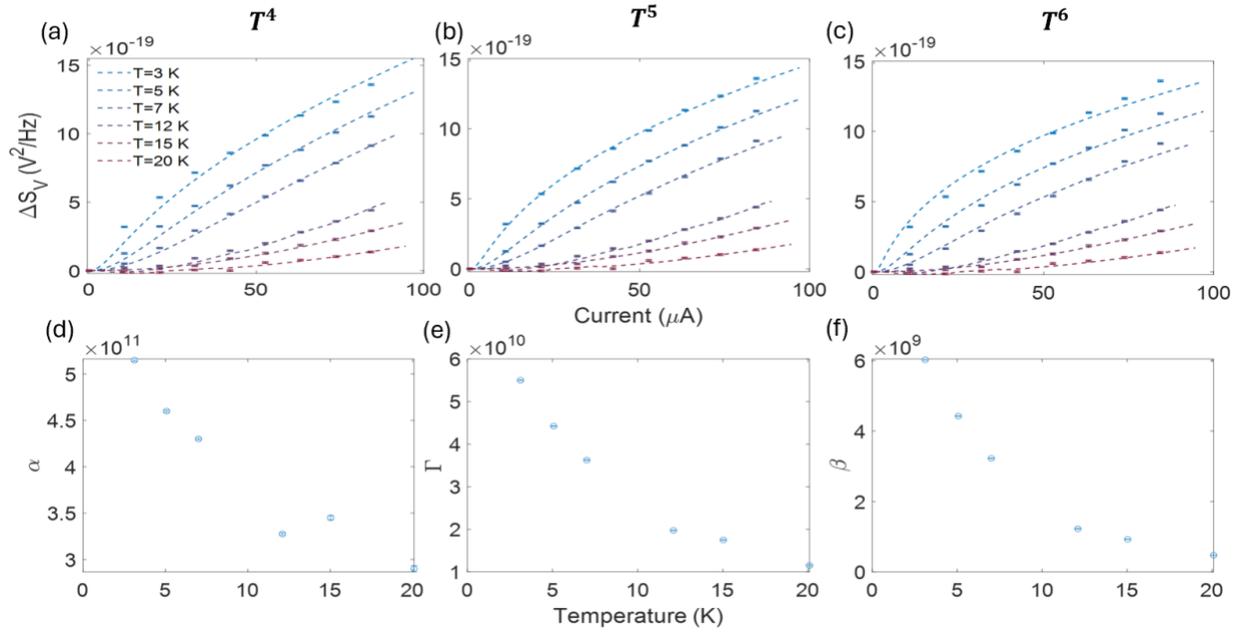

FIG. S3. (a-c) Fits to different versions of Eq. 4 with different temperature dependencies: a) $\alpha(T_e^4 - T_{ph}^4)$, b) $\Gamma(T_e^5 - T_{ph}^5)$, c) $\beta(T_e^6 - T_{ph}^6)$. (d-f) Temperature dependence of the corresponding electron-phonon energy transfer parameters.

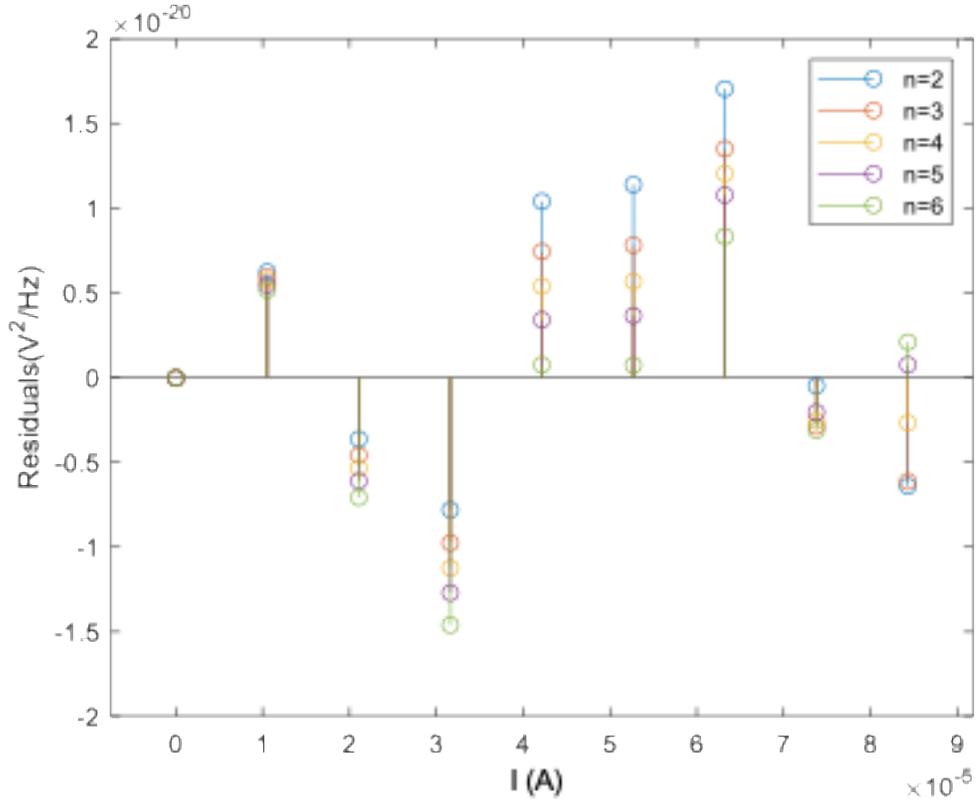

FIG S4: Plot of the residuals of the best fits of each fixed-exponent, single-prefactor-at-each-temperature model to the experimental noise data at the substrate temperature of 15 K.

| Exponent of $T_e^n$-$T_{ph}^n$ term | Variance | $T_e$ |
|---|---|---|
| n=2 | 7.575e-41 | 16.763 |
| n=3 | 5.509e-41 | 16.747 |
| n=4 | 4.520e-41 | 16.731 |
| n=5 | 4.158e-41 | 16.714 |
| n=6 | 4.163e-41 | 16.697 |

TABLE S2: List of different models using different values of $n$ for the exponents in the $T_e^n$-$T_{ph}^n$ and the corresponding variance ($V^4/Hz^2$) over the whole bias range and electron temperature (K) of each fixed-exponent model at the highest bias (85 µA) for the substrate temperature 15 K.

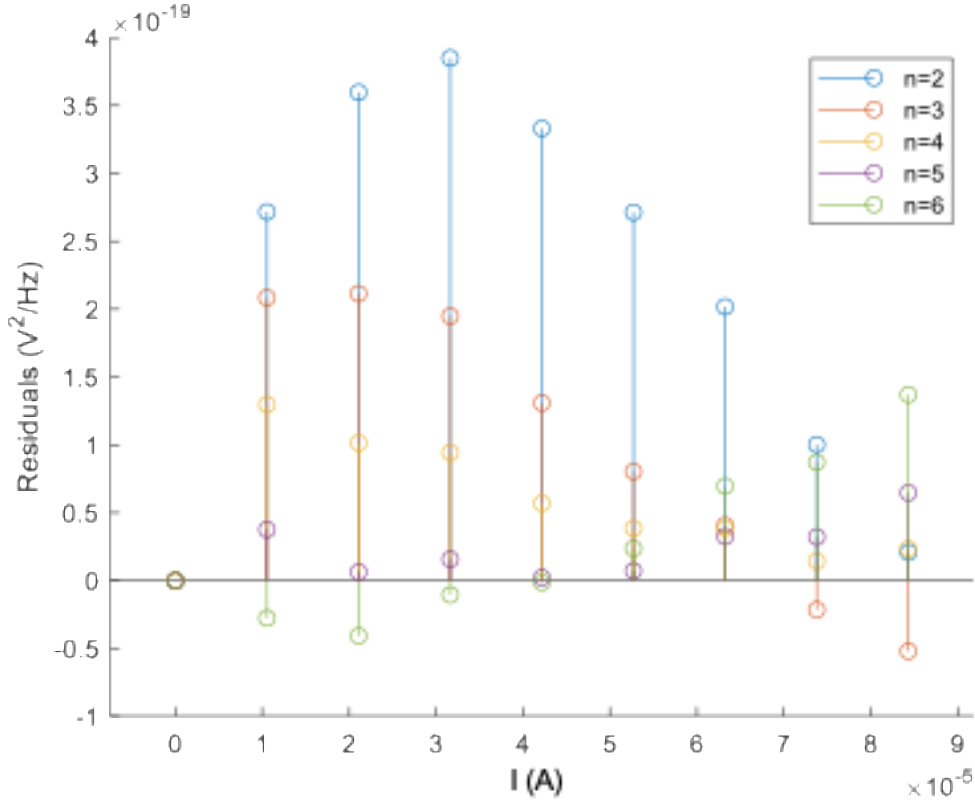

FIG S5: Plot of the residuals of the best fits of each fixed-exponent, single-prefactor-at-each-temperature model to the experimental noise data at the substrate temperature of 3 K.

| Exponent of $T_e^n$-$T_{ph}^n$ term | Variance | $T_e$ |
|---|---|---|
| n=2 | 6.525e-38 | 12.985 |
| n=3 | 1.717e-38 | 12.187 |
| n=4 | 4.768e-39 | 11.563 |
| n=5 | 8.908e-40 | 11.076 |
| n=6 | 3.811e-39 | 10.722 |

TABLE S3: List of different models using different values of $n$ for the exponents in the $T_e^n$-$T_{ph}^n$ and the corresponding variance ($V^4/Hz^2$) over the whole bias range and the electron temperature (K) of each fixed-exponent model at the highest bias (84 µA) for the substrate temperature 3 K.

***Differential resistance.*** The differential resistance under bias is used for analyzing the noise measurements via Eq. (4) of the main text. The wires in this case are quite Ohmic (see Fig. S6), so this is a minor point that does not significantly affect the inferred electron temperatures or the inferred electron-phonon energy loss parameter $\Gamma$.

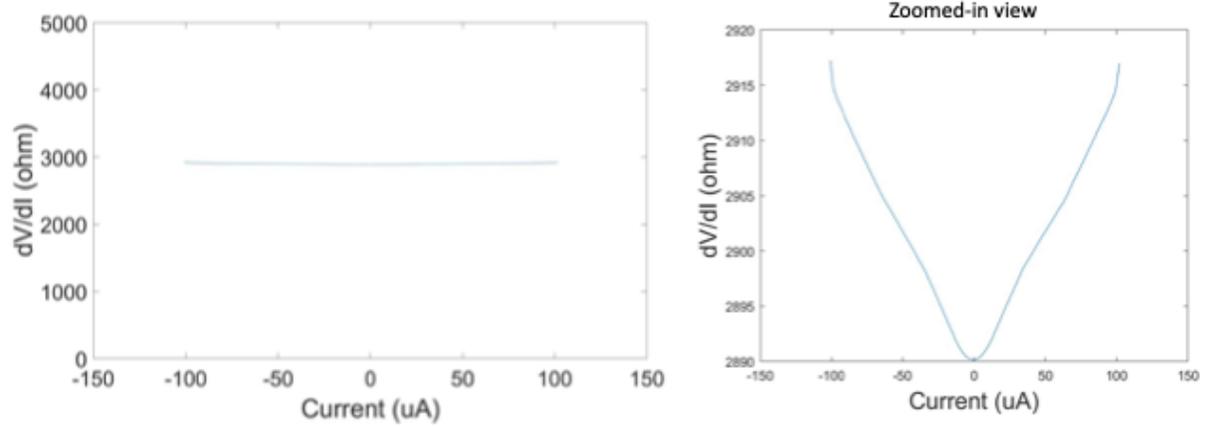

FIG. S6. Differential resistance of the wire used for the noise measurements in Fig. 3 as a function of bias current. Right panel: A zoomed-in view, showing that dV/dI varies from zero bias to the maximum current by a little less than 1%.

*Additional electronic structure details.* Building on the DFT+DMFT calculations, for YbAl$_3$ in a cubic environment, the $J = 7/2$ multiplet splits into $\Gamma^6 \oplus \Gamma^7 \oplus \Gamma^8$. Based on the charge self-consistent DFT+DMFT results, we find that the energy difference between the doublet ground state $\Gamma^6$ and the first excited quartet state $\Gamma^8$ is about 3 meV, equivalently 34.8 K, which is very close to the coherence temperature of ∼ 37 K). This is suggestive that the onset of coherent heavy quasiparticles coincides with the reduction to an effective spin-1/2 manifold appropriate for the Kondo lattice picture of heavy fermions. This is also suggestive of a mechanism for the onset of the roughly linear-in-$B$ magnetoresistance background in Fig. 1b. While very challenging to model directly, Zeeman splitting of the ground state doublet (linear in $B$-field) could then lead to a corresponding reduction in hybridization, suppression of $g(\epsilon_F)$, and a corresponding increase in residual resistivity at fixed disorder, an idea that could be tested with future calculations. The lack of dependence of the linear-in-$B$ magnetoresistive background (Fig. 1b) the field direction is consistent with this Zeeman picture, assuming an isotropic $g$-factor.

The energy difference between the second excited doublet $\Gamma^7$ and the ground state $\Gamma^6$ is approximately 25 meV, which is quite close to optical phonon frequency where Al has the prominent contribution to the phonon DOS. This proximity in energy scale could also lead to stronger interplay of the crystal electric field and the phonons, as proposed in the work of Čermák et al. [7].

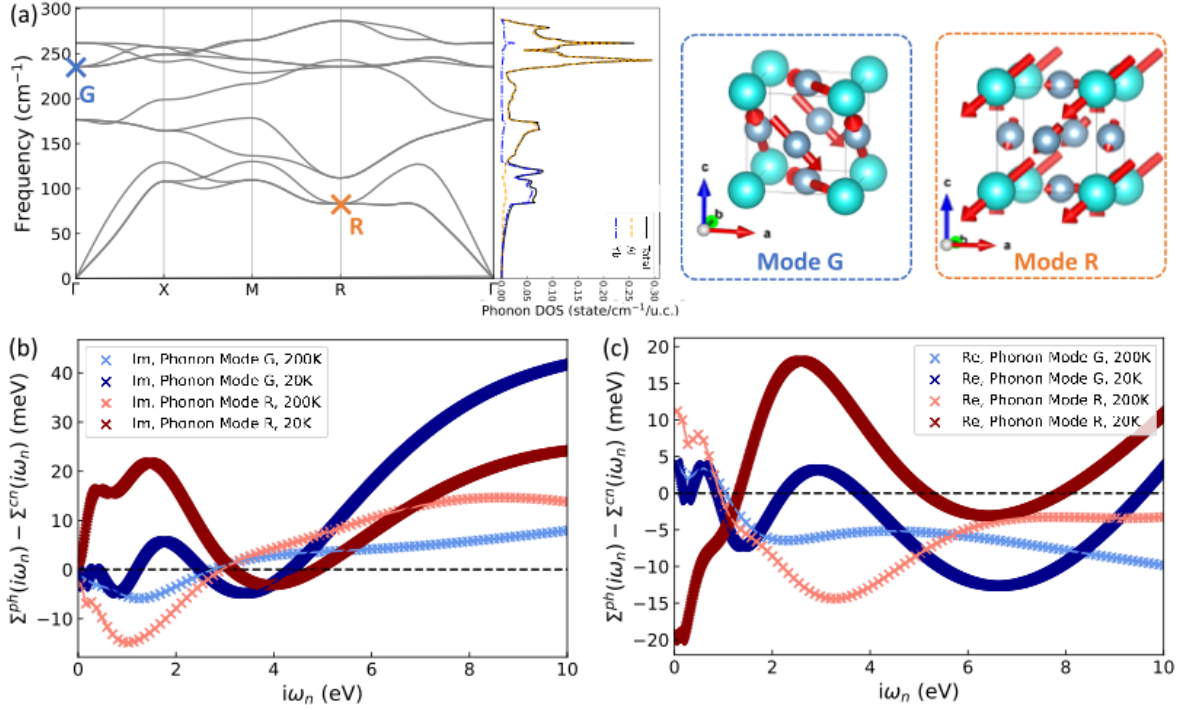

FIG. S7 (color online). (a) Phonon band structure and phonon DOS of YbAl$_3$ with resolved contributions from Al and Yb. The frequencies corresponding to the distortion modes G and R are marked on the phonon band structure, with the atomic displacements illustrated on the right panel. The imaginary (b) and real parts (c) of self-energy difference ($\Sigma^{ph} - \Sigma^{cn}$) corresponding to J = 7/2 between the perturbed cell ($\Sigma^{ph}$) and the unperturbed cell ($\Sigma^{cn}$) at T = 20 and 200 K for modes G and R.

In the DFT calculations, the Kohn-Sham orbitals were solved using the WIEN2k package implementing a full-potential linear augmented plane-wave formalism [8]. The continuous time quantum Monte Carlo (CTQMC) impurity solver was used by considering only a finite number of valences of Yb, including $4f^{11}$, $4f^{12}$, $4f^{13}$, and $4f^{14}$ configurations. A nominal double counting scheme was employed, which was verified to be very close to the exact double counting [9]. Note that for the low-temperature cases, in each full DFT+DMFT loop, two DMFT loops were adopted to ensure good convergence. The DOS was then calculated by implementing the analytical continuations on the Matsubara self-energy functions [10] after the fully converged DFT + DMFT calculations at T = 17.5, 20, 30, 50, 100, and 200 K.

In the phonon-related calculations, Yb $4f$ states were treated as valence states and a DFT + U method was used, with the value of U having a rather limited effect on the phonon band structure [11].